# *Optimization of parameters of nanostructure for study inverse proximity effects on "superconductor-ferromagnetic" interface using Polarized Neutron Reflectometry in enhanced standing wave regime.*

Yu. N. Khaidukov[1], Yu.V. Nikitenko [1], V.L. Aksenov [2]
[1] Frank Laboratory of Neutron Physics, Joint Institute for Nuclear Research, Dubna, Russia,
[2] Russian Research Centre "Kurchatov Institute", Moscow, Russia)

Abstract
This work is devoted to experimental study of influence of superconductivity (S) on ferromagnetism (FM) (inverse proximity effects) with the help of Polarized Neutron Reflectivity. Combining meausurements of specular and diffuse intensities it is possible to obtain full picture of magnetization change in S/FM layered systems like magnetization rotation, domain state formation, inducing of magnetization in S layer, etc. To increase weak magnetic signal we propose to use enhanced neutron standing wave regime (e.g. waveguides). Choose of materials, optimization of thicknesses of layers, estimation of roughnesses influence is presented in this work.

## *Introduction.*

Influence of superconductivity (S) and ferromagnetism (FM) is a very interesting phenomenon both for fundamental science and practical applications. It is predicted that due to proximity between S and FM layers there may be different scenarios of influence of superconductivity on ferromagnetism as domain formation (cryptoferromagnetism) in FM layer [1], magnetization leakage from FM into S layer [2,3], change of indirect exchange coupling of neighboring FM layers through S layer [4], etc. Practical importance of such systems is connected with perspective of creation of compact storage device with two recording channels: on electrical resistance and magnetic order.

One need to say that there is a very small number of performed experiments where influence of superconductivity on magnetism was observed. We can cite two works [5,6] where reduction of magnetization below Tc in S/FM bilayer was detected. The authors of these works said that one of the possible reasons of this effect can be cryptoferromagnetic state. On the other hand the same results can be also explained by effect of magnetization leakage, or so called inverse proximity effects. The essence of this effect is that due to proximity between S and FM layers induced negative magnetization in S layer and suppression of magnetization at FM layer takes place. To choose between these to explanations in this work it is suggested to use Polarized Neutron Reflectometry (PNR). PNR is sensitive to magnetic profile of the system $\vec{M}(\vec{r})$ In case of appearing of domain state in FM layer increase of diffuse scattering [7] will take place, while magnetization leakage will only change the form of specular reflectivity. Our preliminary calculations showed that these changes, especially diffuse scattering are quite small. To increase diffuse and magnetic signal (for example spin flip scattering) we propose to use structures with resonant enhanced neutron standing waves (so called waveguide structures) [8,9].Note that using of regime of resonant enhanced standing waves to increase different secondary process already have been investigated both in X-ray [10,11], and neutron reflectometry [12,13,14]

The goal of this work is optimization of different parameters for creation resonant nanostructure to study weak proximity effects in single S/FM interface.

## *Choice of materials.*

To create resonant structure the investigated S/FM interface must be placed between two thick layers with high optical potential ρ>> 0. Besides this optical condition it is necessary to consider physical aspects of mutual influence of superconductivity and magnetism and technologically aspects of films preparation. In the first case it is required to create such S/FM structure, where strong superconductor (great values of critical temperature and magnetic field Tc and Hc ) is situated near weak ferromagnet (small Curie temperature Tk). To lower Curie temperature one should use thin FM layers (the order of several monolayers), dilution by non magnetic material, etc. To have a strong superconductor, its thickness should be more some threshold $d_S^{min}$ . Moreover, as a part of resonant structure, the optical potential of a superconductor should be as small as possible.

As to technological aspects it is can be said that for preparation of high-quality structure, it is necessary, that structural parameters of materials (type of a lattice, parameters of elementary cells, etc.) were close to each other. One should take into account, that various technologies of preparation impose the restrictions on thickness of layers. For example, at magnetron sputtering of iron with thickness more than 20Å in Fe/V layered systems the three-dimensional growth of layers takes place, which spoils quality of layers. All the above described properties of some superconducting and ferromagnetic materials are presented in tables 1 and 2.

Table №1. Superconducting, structural properties and neutron scattering length density of some superconducting materials.

| material, | Tc,K | Hcm, Oe | type of lattice | cells parameters a/b/c (A), α/β/γ (grad) | ρ, $10^{-6}$ $A^{-2}$ |
|---|---|---|---|---|---|
| In | 3.41 | 281.5 | tetragonal | 3.3/3.3/5.0, 90/90/90 | 1.55 – 0.2i |
| La(alfa) | 4.88 | 800 | hcp | 3.8/3.8/12.1, 90/90/120 | 2.20 |
| La(beta) | 6.00 | 1096,1600 | -- | -- | 2.20 |
| Nb | 9.25 | 2060 | bcc | 3.3/3.3/3.3 90/90/90 | 3.947 |
| Pb | 7.196 | 803 | ccp | 4.9/4.9/4.9 90/90/90 | 3.103 |
| Sn | 3.722 | 305 | tetragonal | 5.8/5.8/3.2 90/90/90 | 2.302 |
| Tc | 7.8 | 1410 | hcp | 2.7/2.7/4.4 90/90/120 | 4.8 |
| V | 5.4 | 1408 | bcc | 3.0/3.0/3.0 90/90/90 | -0.27 |

Table №2. Ferromagnetic, structural properties and neutron scattering length density of some ferromagnetic materials.

| material | Tк,K | Bsat(@RT), Oe | type of lattice | cells parameters a/b/c (A), α/β/γ (град) | ρ, $10^{-6}$ $A^{-2}$ |
|---|---|---|---|---|---|
| Fe | 1044 | 21600 | bcc | 2.9/2.9/2.9 90/90/90 | 8.02 |
| Co | 1388 | 18200 | hcp | 2.5/2.5/4.0 90/90/120 | 2.26 |
| Ni | 627.4 | 6200 | ccp | 3.5/3.5/3.5 90/90/90 | 9.41 |

| | | | | | |
|---|---|---|---|---|---|

Considering all aforesaid, it is offered to create the neutron-resonant structure consisting of thick (~1 mm) MgO substrate ($\rho=6\text{x}10^{-6}$ Å$^{-2}$) on which the layer of iron $^{57}$Fe with thickness of the order of 2-4 monolayers is put , then a layer of a superconductor with thickness $d_S$ and then a layer of high-reflective copper ($\rho=6.6\text{x}10^{-6}$ Å$^{-2}$) with thickness $d_l$. It is suggested to use 57 isotope of iron in order to have possibility to make complementary experiment with Mossbauer spectrometry. If necessary it is possible to replace it by any material presented in table 2. The nuclear profile of structure is presented on fig.1

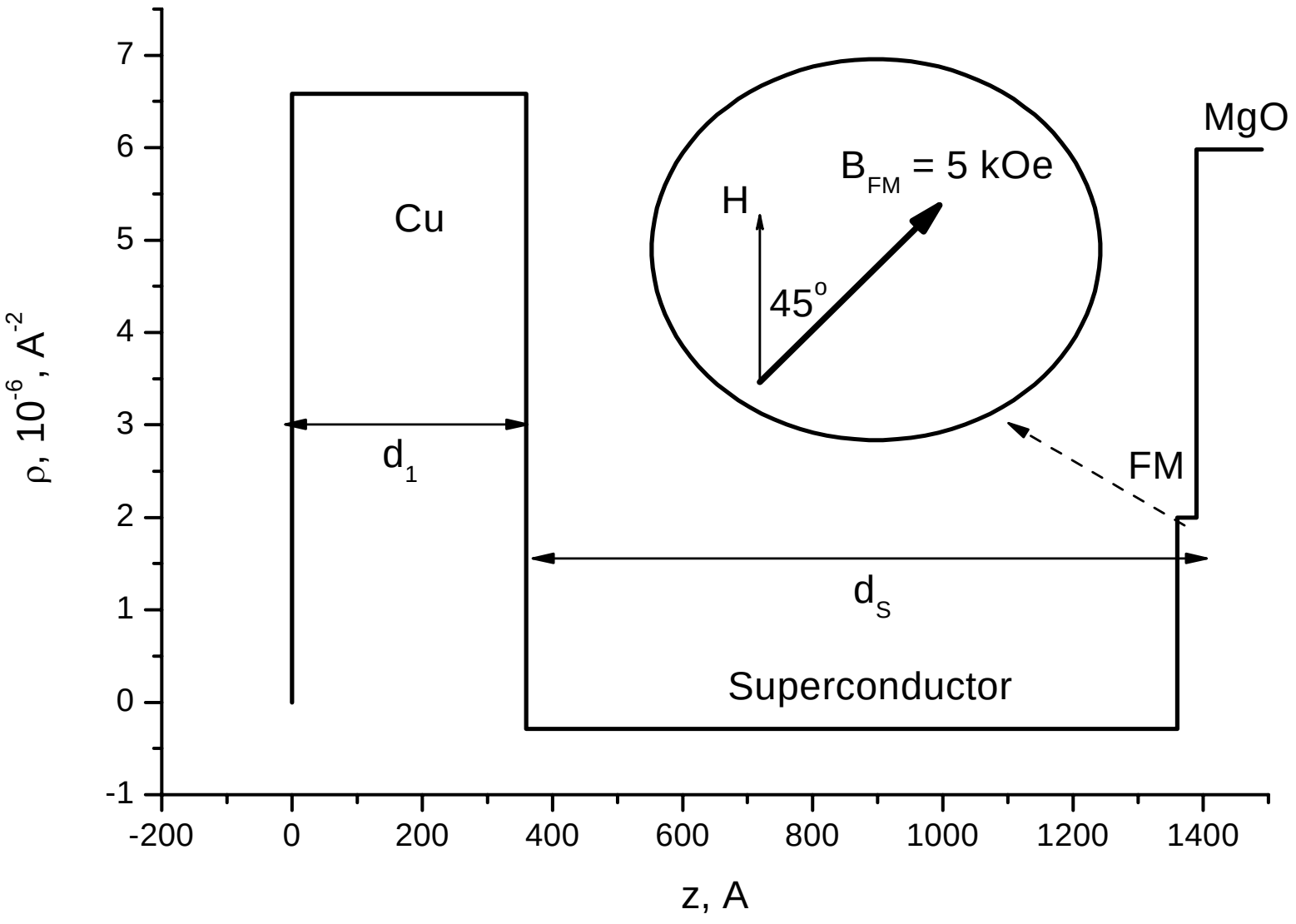

Fig.1. Nuclear profile of suggested resonant structure Cu($d_l$)/S($d_S$)/Fe(2ML)/MgO

## *Optimization of thicknesses.*

From the superconductors presented in table №1 the best candidates by criterion «the maximal Tc at minimal ρ» are Nb, Pb and V. Dependences Tc ($d_S$), taken from [15], [16] and [17] for the given materials are presented on fig.2. Using these data we limit from below thickness for niobium, lead and vanadium as $d_S^{min}$ =370Å, $d_S^{min}$ =630Å and $d_S^{min}$ =400Å accordingly.

Optimization of thickness $d_l$ and $d_S$ with the purpose to obtain the maximal secondary signal (spin flip and diffuse scattering) was done for these materials. All calculations were done at the final resolution of experiment dQ/Q=1.5 %. Behaviour of diffuse and spin-flip scattering turned out to be very similar for such a kind of optimizations. That's why all presented here calculations were done for spin-flip as secondary process. On fig. 3 maximum of spin flip $R^{+-}(Q)$ as a function of $d_l$ and $d_S$ for chosen superconductors is presented. It can be seen that several sets of thicknesses can be used to create resonators (table 3).

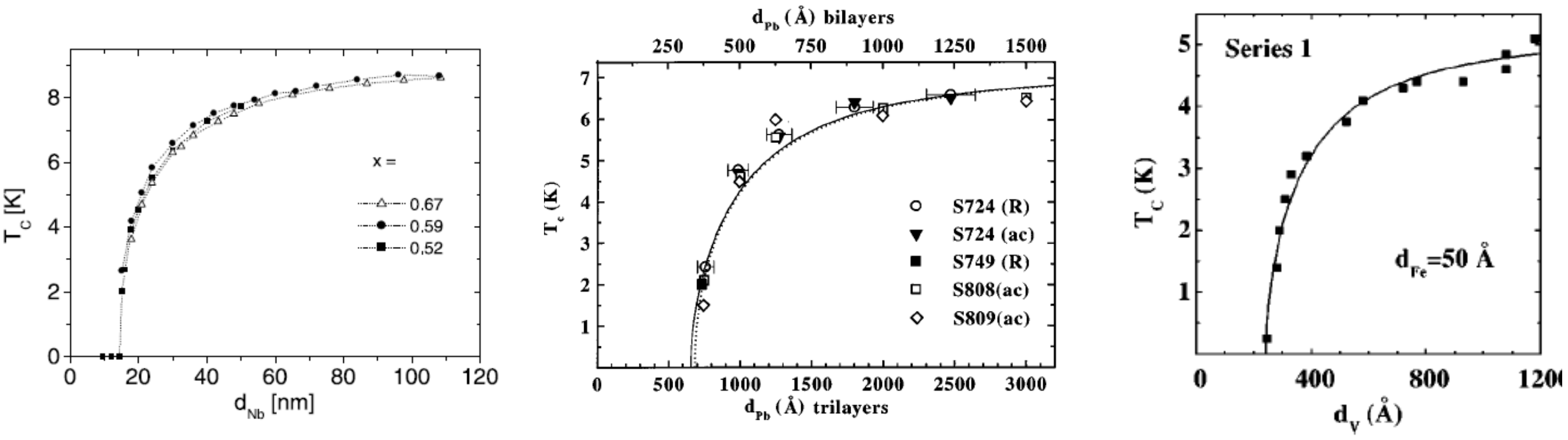

Fig. 2 Dependence of the Tc on thickness of a superconductor for Nb [6], Pb [7] and V [8]

a)

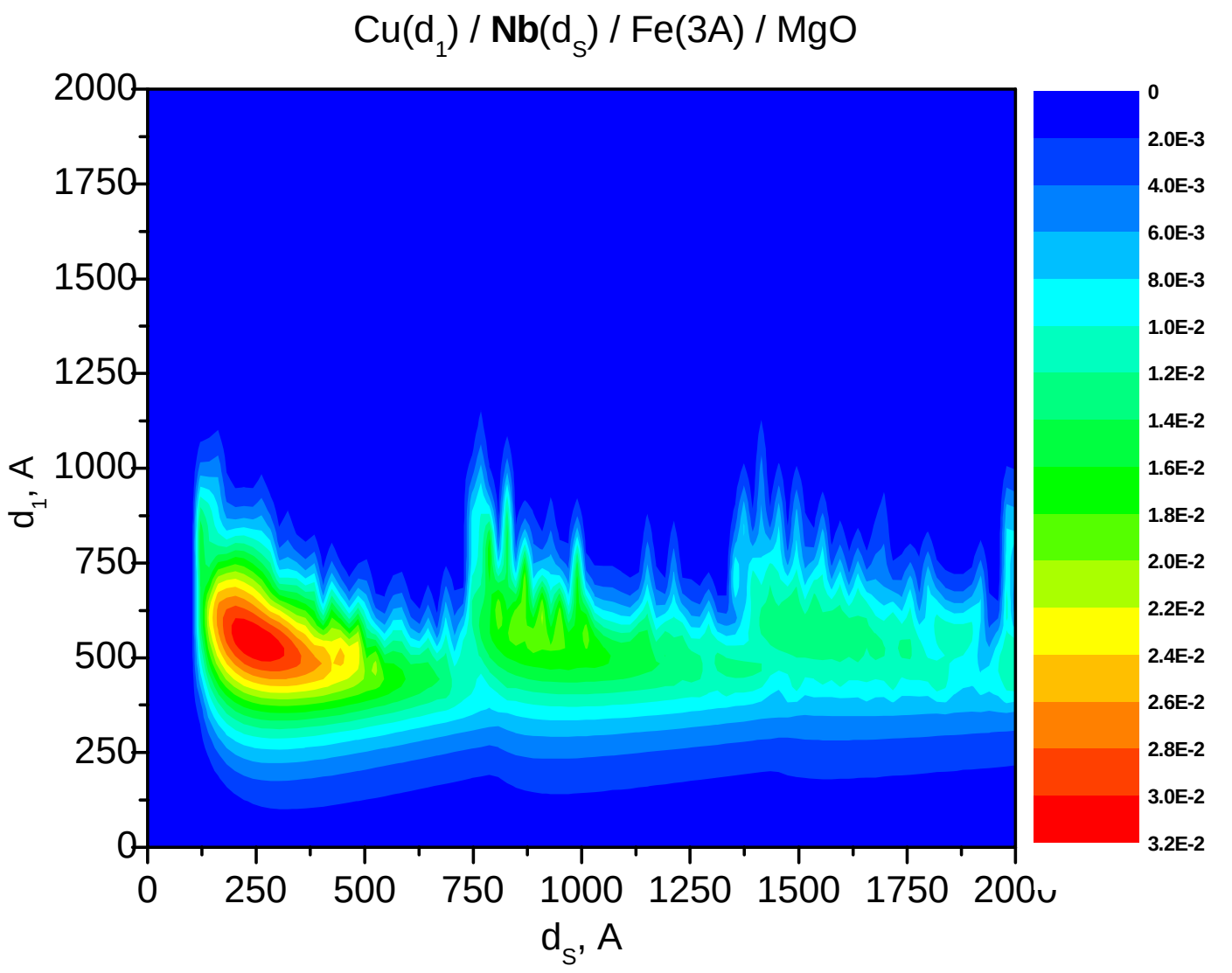

b)

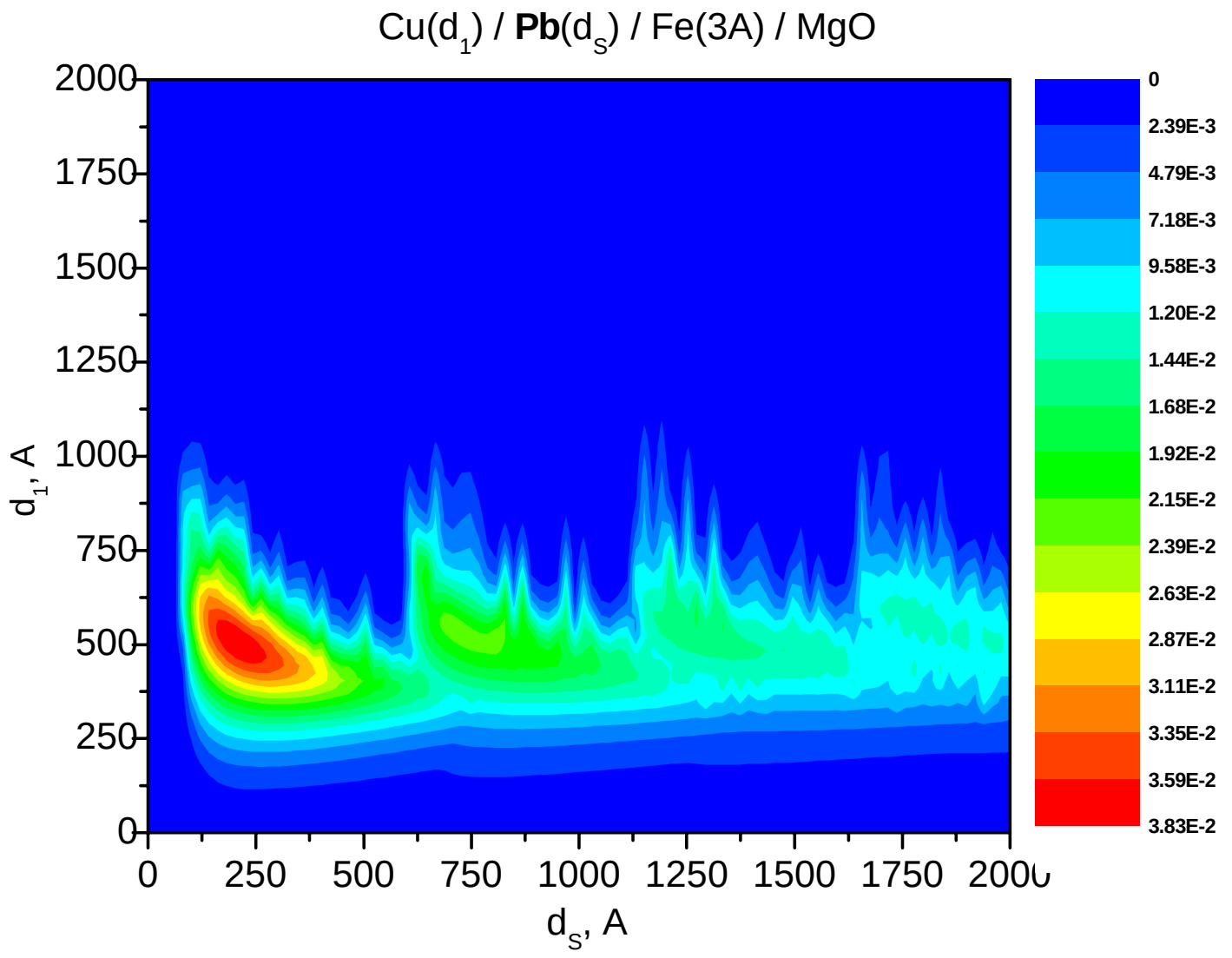

c)

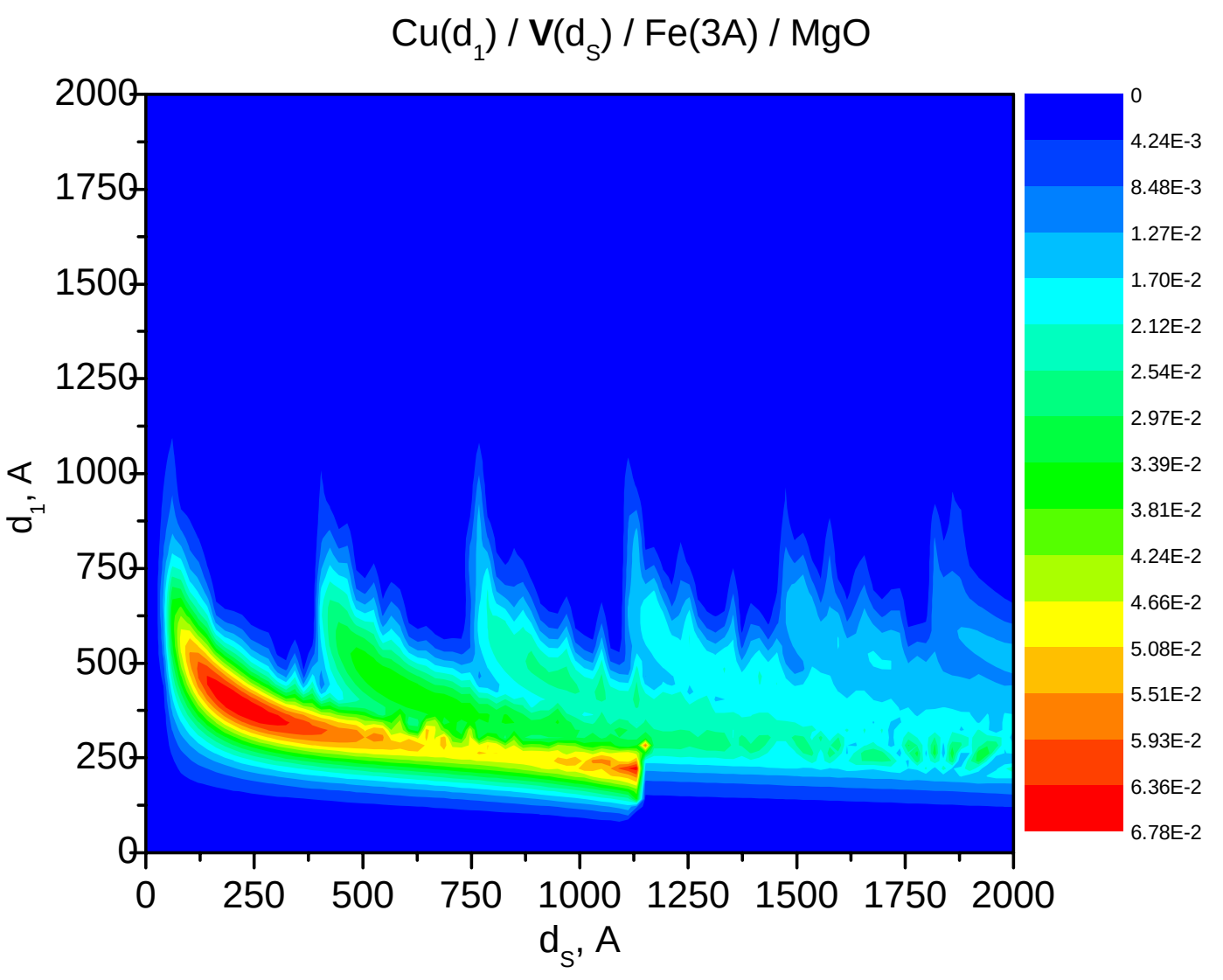

Fig. 3 Dependence of maximum of $R^{+-}(Q)$ spin-flip signal on thickness $d_1$ and $d_S$ for chosen superconductors

Table №3. Optimum thickness of layers of copper and a superconducting material

| superconductor | $d_S^{min}$, Å | $d_1$, Å | $d_S$, Å | maximum spin-flip signal, % |
|---|---|---|---|---|
| niobium | 370 | 400-570 | 370-490 | 2-2.5 |
| niobium | 370 | 485-680 | 820-1090 | 1.5-2 |
| lead | 625 | 440-600 | 680-950 | 2 |
| vanadium | 400 | 280-350 | 400-1100 | 5-7 |

## *Influence of roughness.*

At preparation of real structures occurrence of roughnesses is inevitably. The given roughnesses can influence on structures doubly. First, presence of roughnesses leads to degradation of interfaces that spoils resonator properties of structure. Calculations on influence of the given effect were done for the optimized structures from table 3.. On fig. 4 the maximum spin flip signal as function of rms height of roughnesses on interfaces "vacuum/copper" ($\sigma_1$) and "copper/vanadium" ($\sigma_2$) for structure Cu (546 Å)/V (242 Å)/$^5$7Fe (3-6 Å)/MgO is presented. It is visible, that resonant enhancement survives at roughnesses $\sigma_{1(2)} \leq$ 100Å. Similar calculations for niobium and lead show, that maximum permissible roughnesses for these materials less and make ~30 Å. In view of the calculations done for other interfaces, we estimate maximum allowed value of roughness at given interface as 20-30Å.

Besides this it is necessary to consider, that presence of roughnesses on S/FM interface leads to penetration of non- magnetic atoms in FM layer that leads to downturn of concentration of magnetic atoms and as consequence, temperatures Tk [15] and average magnetization [18].

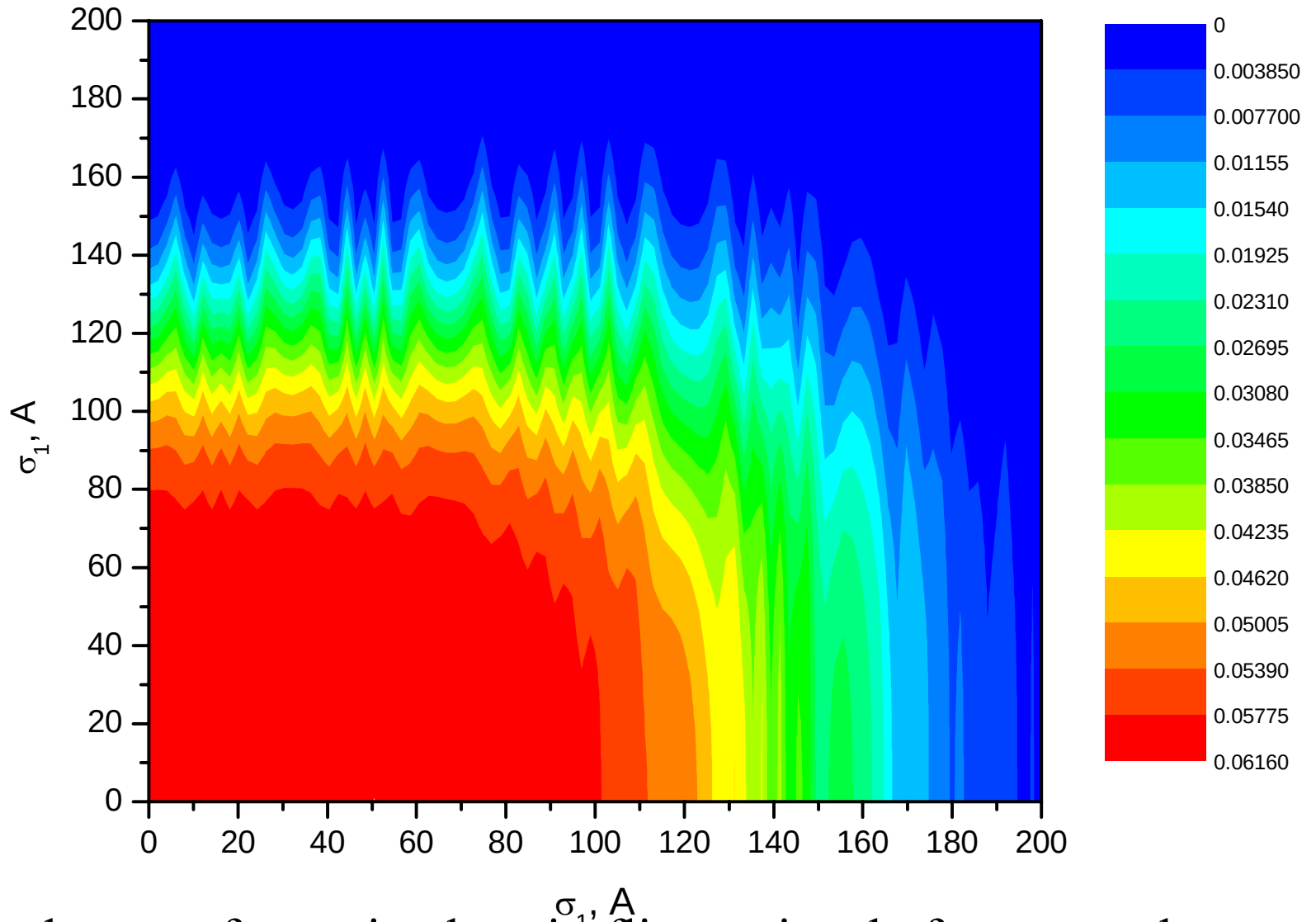

Fig. 4. Dependence of maximal spin-flip signal from roughnesses on interfaces "vacuum/copper" $\sigma_1$ and "copper/vanadium" $\sigma_2$ for structure Cu(546 Å)/V(400 Å)/$^{57}$Fe(3-6Å)/MgO. 6Å)/MgO.

## *Estimation of effect.*

On fig. 5 spin flip reflectivity $R^{+-}(Q)$ for optimized structure Cu(323Å)/V(400Å)/Fe(3Å)/MgO is presented. Resonant enhancement is seen at Q=0.001 Å$^{-1}$. In the same figure spin flip reflectivity for structure without copper layer is presented. Factor of enhancement makes $1.3x10^3$. Simple calculations show, that at intensity of neutrons $10^3$ n/sec, such structure will allow to observe changes of the magnetic moment in FM layer of the order of 1 %.

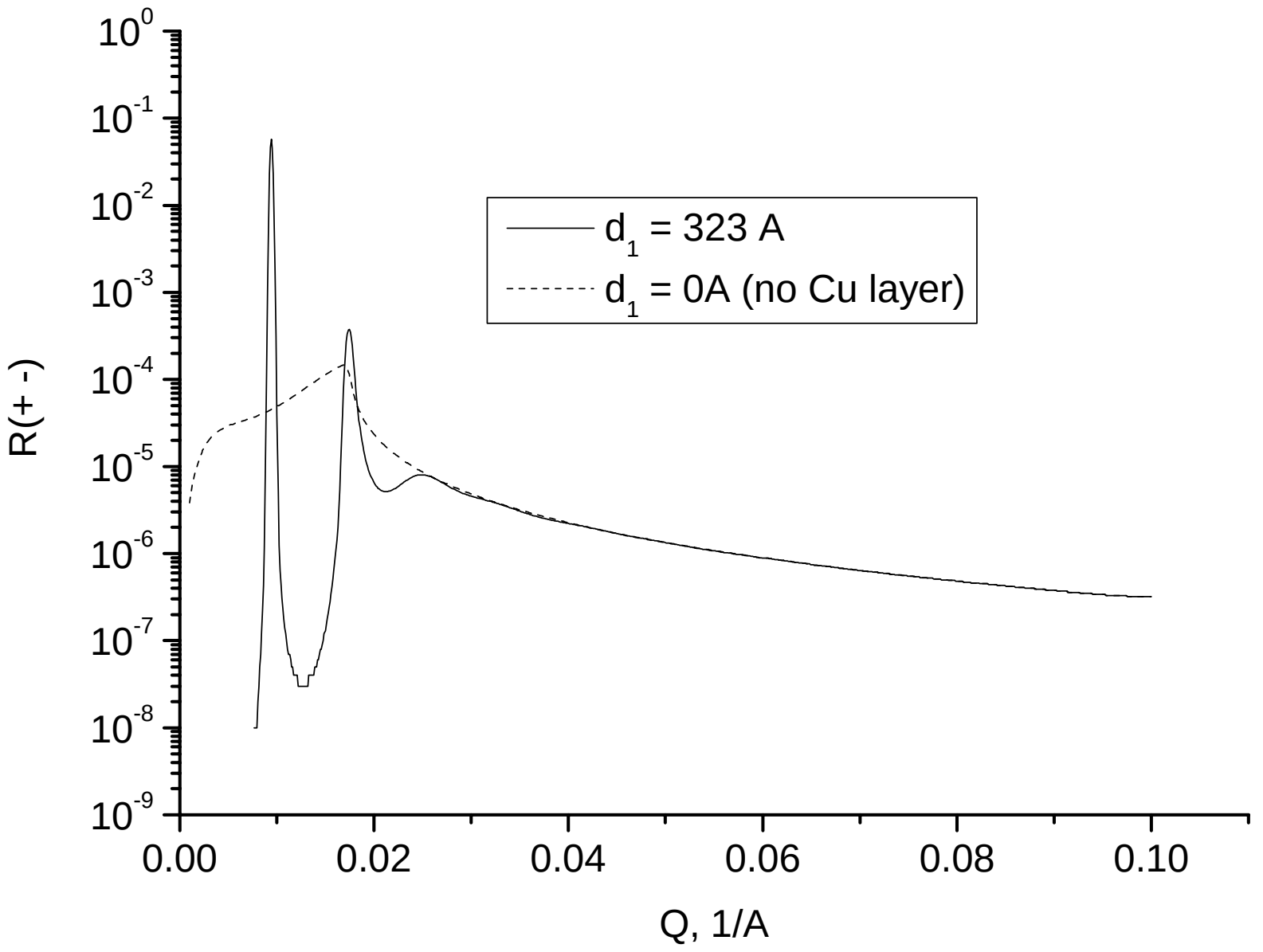

Fig. 5. Spin flip reflectivity for system with a layer of copper (a continuous line) and without a layer (shaded).

## *Conclusions.*

The regime of enhanced standing waves is very sensitive tool for studying of effects of influence of superconductivity on magnetism. In order to obtain resonant nanostructure the optimization of parameters of layered structure in view of neutron optical, superconducting-ferromagnetic and technological aspects is done. Results of optimization are presented in table 3. Calculations show, that at roughnesses of the order 20-30 Å, neutron flux $10^3$ n/sec and resolution of reflectometer dQ/Q=1.5 % such structures will allow to observe change of magnetization in FM layer of the order of 1 %

## *References.*